\documentclass[aip,jmp,reprint,amsmath,amssymb]{revtex4-1}

\usepackage{bbold}
\usepackage{bbm}
\usepackage[pdftex]{graphicx}
\usepackage{graphicx}
\usepackage{latexsym,amsmath,verbatim}
\usepackage{color}
\usepackage{rotating}
\usepackage{multirow}
\usepackage[english]{babel}
\usepackage{color}
\usepackage{changes}
\usepackage{todonotes}
\usepackage{amssymb}

\newcommand{\header}[1]{\noindent\textbf{\emph{#1.}}}
\newcommand{\rev}[1]{{\color{black} #1}}

\begin{document}

%%%%%

\title{\rev{Direct measurement of superdiffusive and subdiffusive energy transport in disordered granular chains}}

\author{Eunho Kim$^{*}$}
\affiliation{Department of Aeronautics and Astronautics, University of Washington, Seattle, WA 98195-2400, USA}
\affiliation{Division of Mechanical System Engineering \& Automotive Hi-Technology Research Center, Chonbuk National University, 567 Baekje-daero, Deokjin-gu, Jeonju-si, Jeollabuk-do, 54896, Republic of Korea}

\author{Alejandro J. Mart\'inez$^{*}$}
\affiliation{Oxford Centre for Industrial and Applied Mathematics, Mathematical Institute, University of Oxford, Oxford OX2 6GG, UK}

\author{Sean E. Phenisee}
\affiliation{Department of Aeronautics and Astronautics, University of Washington, Seattle, WA 98195-2400, USA}

\author{P. G. Kevrekidis}
\affiliation{Department of Mathematics and Statistics, University of Massachusetts, Amherst, Massachusetts 01003-4515, USA}

\author{Mason A. Porter}
\affiliation{Department of Mathematics, University of California, Los Angeles, CA 90095, USA}
\affiliation{Oxford Centre for Industrial and Applied Mathematics, Mathematical Institute, University of Oxford, Oxford OX2 6GG, UK}
\affiliation{CABDyN Complexity Centre, University of Oxford, Oxford OX1 1HP, UK}

\author{Jinkyu Yang}
\affiliation{Department of Aeronautics and Astronautics, University of Washington, Seattle, WA 98195-2400, USA}

\noindent *The first two authors contributed equally. %\\ 
%\noindent Corresponding author: Mason A. Porter 
%\date{}

%%%%%%%%%%%%

%\author{Eunho Kim$^{*1,2}$, Alejandro J. Mart\'inez$^{*3}$, Sean E. Phenisee$^{1}$, P. G. Kevrekidis$^{4}$, Mason A. Porter$^{3,5,6}$, and Jinkyu Yang$^{1}$}

%\maketitle

%\begin{affiliations}
%\item Department of Aeronautics and Astronautics, University of Washington, Seattle, WA 98195-2400, USA
%\item Division of Mechanical System Engineering \& Automotive Hi-Technology Research Center, Chonbuk National University, 567 Baekje-daero, Deokjin-gu, Jeonju-si, Jeollabuk-do, 54896, Republic of Korea
%\item Oxford Centre for Industrial and Applied Mathematics, Mathematical Institute, University of Oxford, Oxford OX2 6GG, UK
%\item Department of Mathematics and Statistics, University of Massachusetts, Amherst, Massachusetts 01003-4515, USA
%\item Department of Mathematics, University of California, Los Angeles, CA 90095, USA
%\item CABDyN Complexity Centre, University of Oxford, Oxford OX1 1HP, UK
%\end{affiliations}
%\noindent *The first two authors contributed equally. \\ 
%\noindent Corresponding author: Mason A. Porter 
%\date{}

%%%%%%%%%%%%%%%%% END OF PREAMBLE %%%%%%%%%%%%%%%%

%\begin{document}
% Double-space the manuscript.
%%%%%\baselineskip24pt
% Make the title.
%\maketitle

%%%%%%

%%%%%%%%%%%%%%%%%%%%%%%%%%%%%%%%%%%%%%%%%%%%%%%%%%%%%%%%%%%%
%%
%%                                                                              A B S T R A C T
%%
%%%%%%%%%%%%%%%%%%%%%%%%%%%%%%%%%%%%%%%%%%%%%%%%%%%%%%%%%%%%

\begin{abstract}

The study of energy transport properties in heterogeneous materials has attracted scientific interest 
for more than a century, and it continues to offer fundamental and rich questions. One of the unanswered challenges is to extend Anderson theory for uncorrelated and fully disordered lattices in condensed-matter systems to physical settings in which additional effects compete with disorder. Specifically, the effect of strong nonlinearity has been largely unexplored experimentally, partly due to the paucity of testbeds that can combine the effect of disorder and nonlinearity in a controllable manner. Here we present the first systematic experimental study of energy transport and localization properties in simultaneously disordered and nonlinear granular crystals. We demonstrate experimentally that disorder and nonlinearity --- which are known from decades of studies to individually favor energy localization --- can in some sense ``cancel each other out'', resulting in the destruction of wave localization. We also report that the combined effect of disorder and nonlinearity can enable the manipulation of energy transport speed in granular crystals from subdiffusive to superdiffusive ranges. 

\end{abstract}

\maketitle

%%%%%%%%%%%%%%%%%%%%%%%%%%%%%%%%%%%%%%%%%%%%%%%%%%%%%%%%%%%%
%%
%%                                                                              I N T R O D U C T I O N 
%%
%%%%%%%%%%%%%%%%%%%%%%%%%%%%%%%%%%%%%%%%%%%%%%%%%%%%%%%%%%%%

\noindent Over the past decade, there have been amazing experimental advances, in fields such as ultracold atomic physics and nonlinear optics, towards the direct observation of spatial localization and transport in
disordered systems\cite{AndersonBECs1,AndersonBECs2,AndersonOptics}.
There has been simultaneous progress towards achieving a theoretical understanding of the interplay between disorder and weak nonlinearity\cite{ref5}. However, there has been much less exploration of disorder in strongly nonlinear systems, and many fundamental questions remain open. Specifically, under what conditions is transport subdiffusive, and more generally how does strong nonlinearity affect localization? Much of the progress has arisen from studies of models
with on-site nonlinearities, such as discrete nonlinear
Schr{\"o}dinger and Klein--Gordon models, where the
interplay between disorder and nonlinearity yields subdiffusive
transport\cite{ref5,flach08,flach09}. However, recent
progress has hinted at a fundamentally different phenomenology
in lattices with inter-site interactions (e.g., well-known settings such as chains of Fermi--Pasta--Ulam (FPU) type\cite{FPU}). In particular, it was shown numerically that superdiffusive behavior is possible in the latter
case\cite{ref4,ref11, ref9}. An intriguing feature in all of the above cases is
that disorder (which is traditionally viewed as leading to localization) and nonlinearity (which can cause localization in the form of phenomena such as discrete breathers\cite{Flach2007}) can somehow cancel out each other's tendency towards localization,
leading to transport.

In this study, we experimentally and numerically investigate energy transport in one-dimensional disordered granular crystals (i.e., granular chains) in a wide range of regimes, extending from nearly linear to strongly nonlinear ones. Granular crystals composed of spherical particles are a popular vehicle for investigating various nonlinear wave features\cite{ref3,sen08,PT2015,chong2016}. When in contact, two particles interact with each other nonlinearly via a Hertzian interaction\cite{ref1}: the force--displacement relation in the contact interaction is governed by a $3/2$ power law under compression and zero force under tension. In this class of intersite-interaction
models, one can tune the effective system nonlinearity very precisely by varying the ratio of excitation amplitude to static precompression\cite{chong2016}. We consider a system excited at a granular chain's boundary to investigate how the mechanical 
energy injected by the external excitation is transported along the chain under the combined influence of nonlinearity and disorder.

%%%%

\vspace{.5cm}

\begin{figure}
\includegraphics
[width=\textwidth]{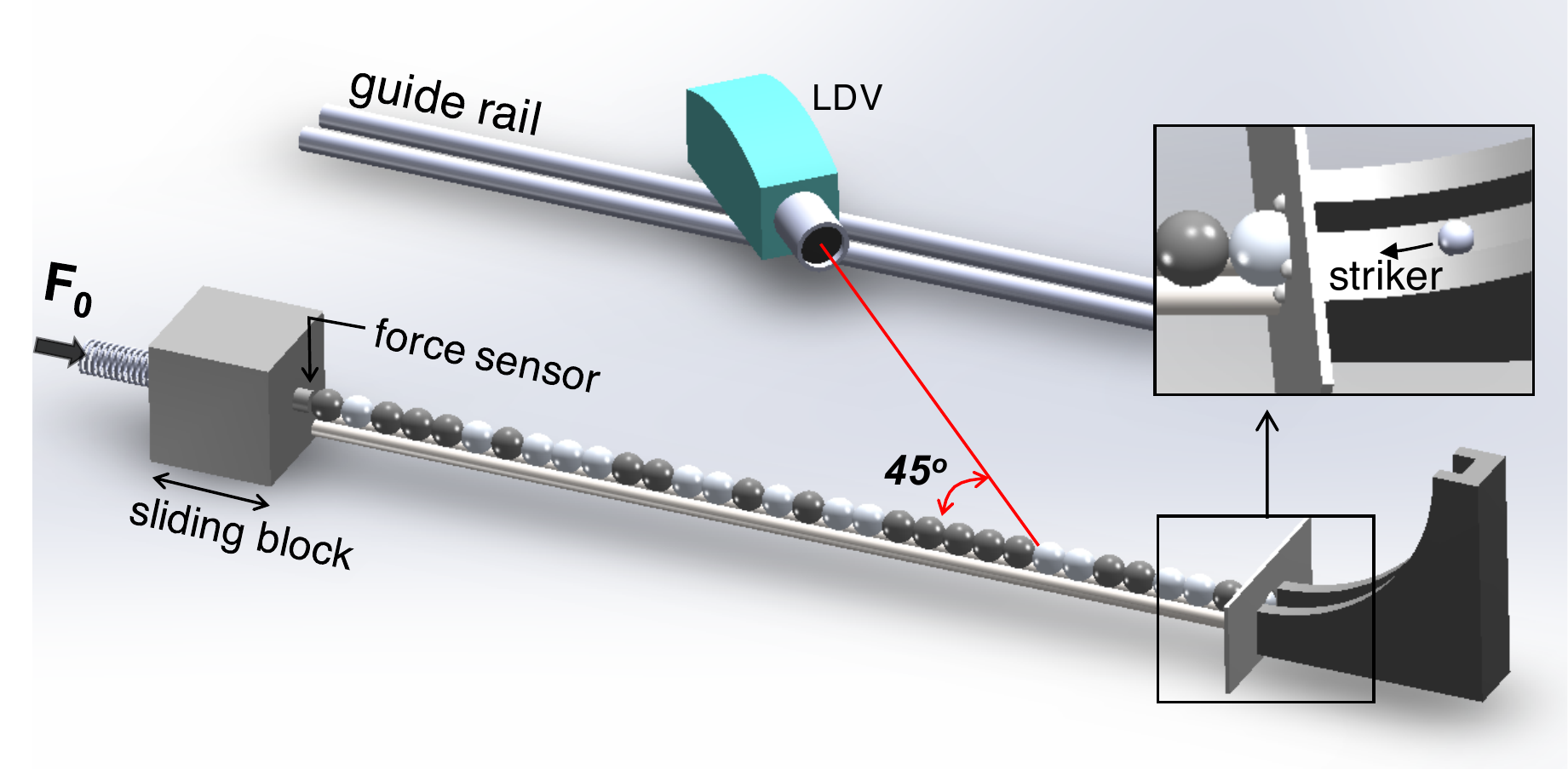}
  \caption{\label{f1} {\bf Schematic of the experimental setup.} The inset shows details of the boundary condition in
    the front of a granular chain. 
}
\end{figure}

\noindent \textbf{Experimental setup and data} \\ \\
\noindent We describe the experimental setup in Figure~\ref{f1}. The granular chain consists of 32 spherical particles, which is long enough to validate energy localization and transport properties in granular crystals (see the Supplemental Information). 
To introduce disorder to a granular crystal, we use various combinations of 
aluminum and tungsten-carbide particles, which have drastic disparities in density and elastic modulus (see Methods for details). 
The right end of the chain is constrained by a steel plate with a hole in the
center; through this hole, a spherical impactor released from a ramp hits and excites the first particle of the chain. The left end of the chain is blocked by a large sliding mass, which applies 
a static precompression $F_0$ to the chain through a linear spring. We measure the dynamics of the chain by recording each particle's velocity as a function of time via a laser Doppler vibrometer (LDV)\cite{ref13}.

%%%%%%%%%%%%%%%%%%%%%%%%%%%%%%%
%%%%%%%%%%%%%%%%%%%%%%%%%%%%%%%

We control the strength of the system's nonlinearity by applying three different levels of precompression: $50$ N, $10$ N, and $0$ N, which range from linear to strongly nonlinear dynamical regimes of the chain. To visualize the propagation of stress waves in each case, we show spatiotemporal distributions of the particles' motions based on the measured velocity. Figure~\ref{f2} gives experiment measurements of particle velocities for a homogeneous chain consisting of aluminum particles only (top row) and a disordered chain composed of aluminum and tungsten-carbide particles (bottom row) in strongly nonlinear, weakly nonlinear, and nearly linear scenarios (left, middle, and right, respectively).

%%%%%%%%%%%%%%%%%%%%%%%%%%%%%%%
\begin{figure}
\includegraphics[width=\textwidth]{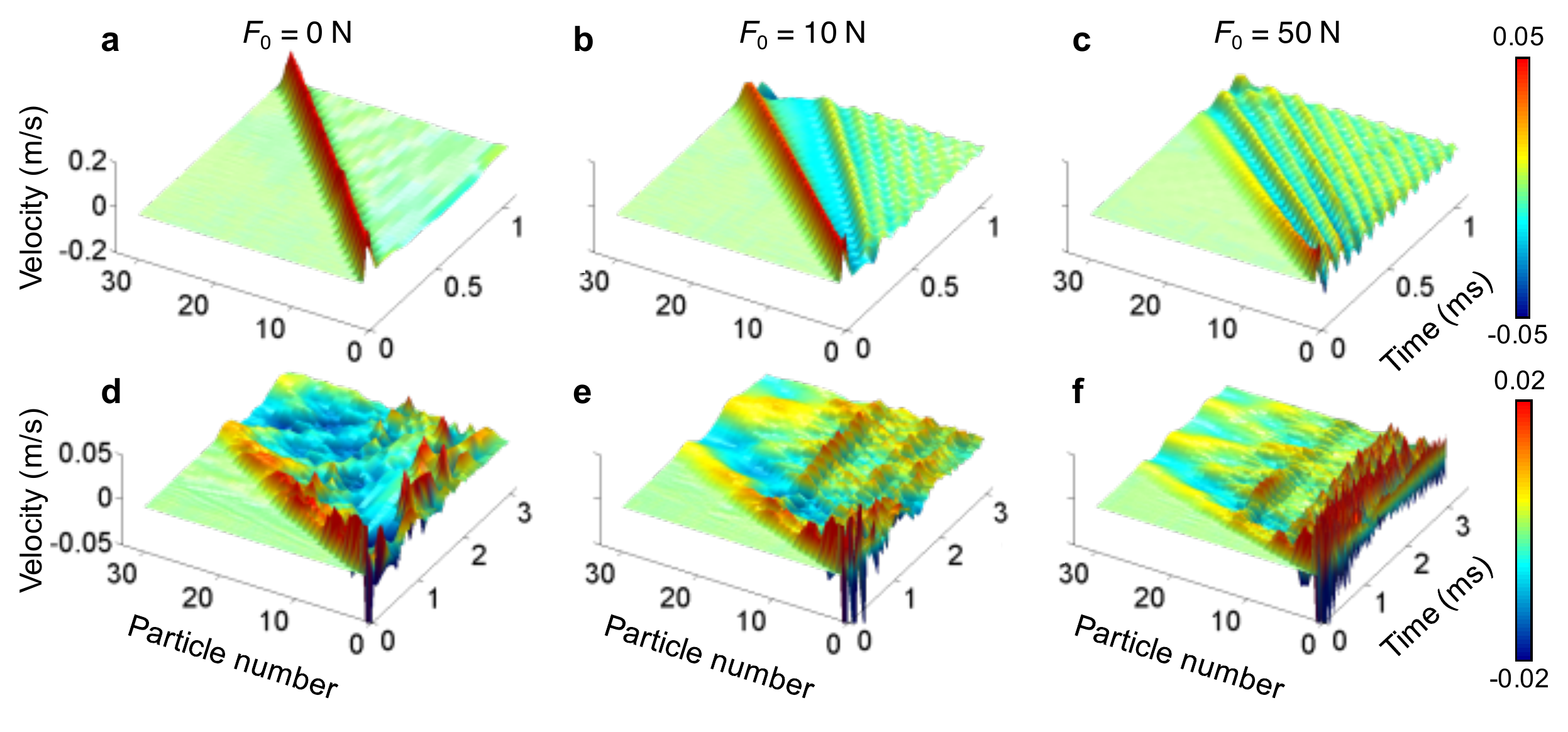}
\caption{\label{f2} {\bf Wave propagation in homogeneous and disordered chains.} Spatiotemporal distributions of particle velocities in ({\bf a--c}) a homogeneous chain and ({\bf d--f}) a disordered chain with static precompressions of ({\bf a, d}) $0$ N, ({\bf b, e}) $10$ N, and ({\bf c, f}) $50$ N.}
\end{figure}

%%%%%%%%%%%%%%%%%%%%%%%%%%%%%%%

In the homogeneous chain, we observe a localized wave packet in the form of a strongly nonlinear solitary wave in the absence of precompression (i.e., for $F_0=0$ N). When we apply a nonzero precompression to the chain, we start
to observe the generation of linear oscillatory waves, which propagate behind the
supersonic leading nonlinear wave packet. These ripples arise from oscillations of the first particle after an impact\cite{ref2} and consequently from the excitation of oscillatory modes. The frequency of these
oscillatory waves increases as the precompression increases (compare Figures~\ref{f2}b and~\ref{f2}c), as the contact stiffness increases with strong precompression because of the nonlinearity in the contact interaction. 

In Figure~\ref{f2}d--f, we show experimental results for wave propagation in a disordered chain for various precompression strengths (See the Supplemental Information for details of the disordered chain configuration, the comparison with the simulation results, and the corresponding frequency analysis). Comparing our results for disordered chains with the ones from the homogeneous chain, we find that the presence of disorder causes significant scattering of propagating waves in both time and space. The scattering is most drastic in the absence of precompression (see Figure~\ref{f2}d). However, for increased precompression, the wave packet tends to become more localized in the front of the chain and the amplitude of propagating waves decrease significantly as a function of 
distance (see Figures~\ref{f2}f). Given identical excitation conditions, we observe that applying precompression increases the speed of the leading wave packet, as expected in granular chains due to the dependence of the wave speed on wave amplitude and chain stiffness (i.e., precompression)\cite{ref3,sen08,PT2015,chong2016}. 

%%%%%%

\vspace{.5cm}

\begin{figure}
\includegraphics[width=\textwidth]{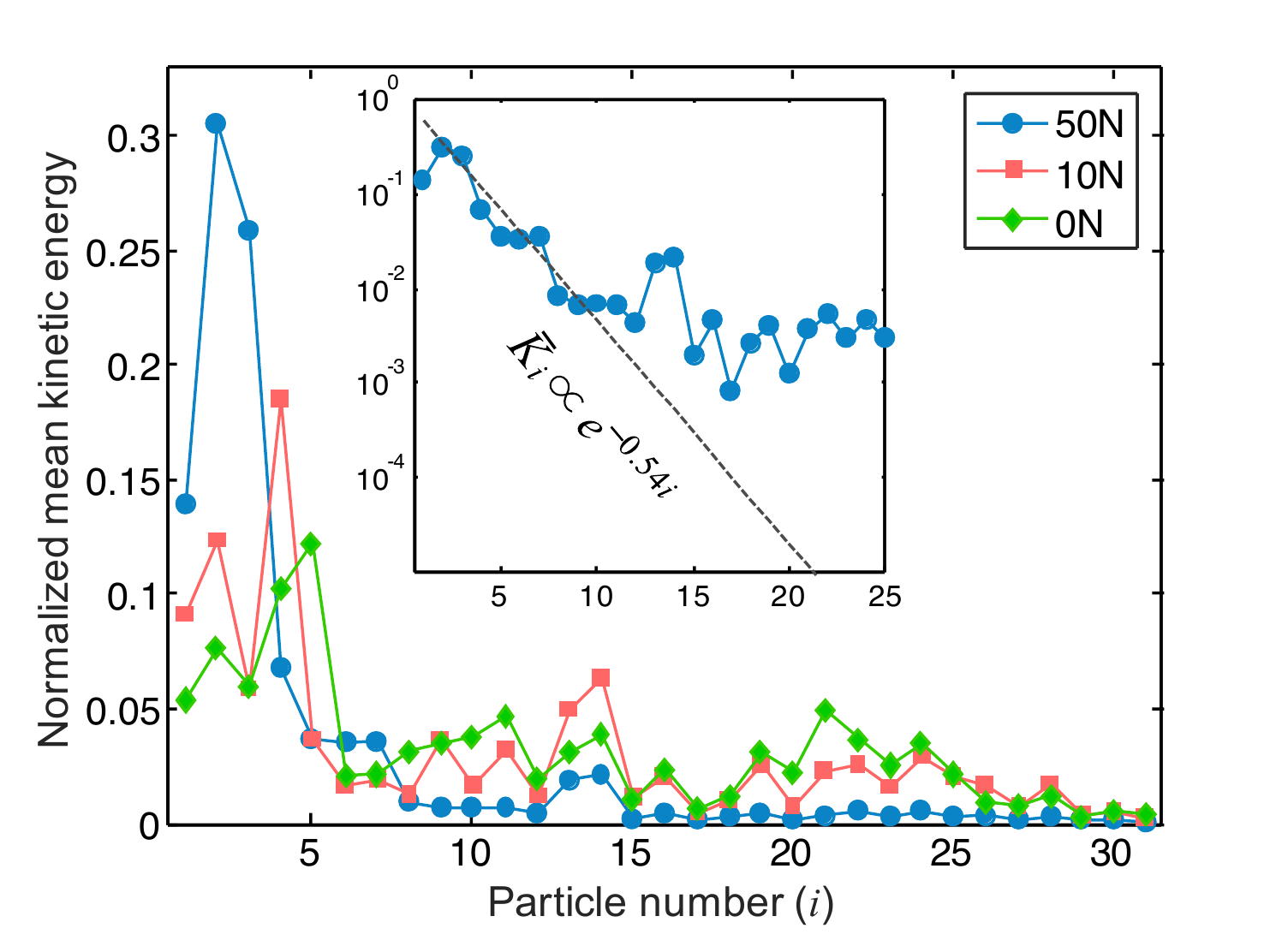}
\caption{\label{f3} {\bf Experimental observation of Anderson-like localization.} Normalized kinetic energy profile, averaged between $1.5$ ms and $3.5$ ms, for different levels 
of precompression. In the inset, we show the kinetic energy on a logarithmic scale when the static precompression is $50$ N. The dashed line shows the slope associated with $e^{-0.54 i}$, where $i$ denotes particle number.}
\end{figure}

\noindent \textbf{Experimental observation of Anderson-like localization} \\ \\
\noindent To characterize the localization phenomenon near the excitation point in the linear regime (i.e., $F_0=50$ N), we compute the kinetic-energy distribution.
Given the velocity $v_i(t)$ of the $i$th particle, we compute its kinetic energy $K_i(t)=(1/2)m_i v_i^2(t)$, where $m_i$ is the particle's mass. We then average our results over the different realizations of disorder, obtaining 
$\left\langle K_i(t)\right\rangle$. Initially, the pattern's amplitude decreases due to the spreading associated with non-scattered modes. It then oscillates near the edge of the chain for a long time (up to $3.5$ ms). 
To better visualize the kinetic-energy profile, we compute a temporal mean between $t_s = 1.5$ ms and $t_f = 3.5$ ms. We define the mean kinetic energy as $\bar{K}_i = \int_{t_s}^{t_f}\left\langle K_i(t)\right\rangle dt/(t_f-t_s)$, 
and we then normalize the distribution by letting $\bar{K}_i\rightarrow \bar{K}_i/\sum_{j=1}^{N}\bar{K}_j$. In Figure~\ref{f3}, we show normalized mean kinetic energy profiles for the different precompression strengths. 
In our experiments, we observe for $F_0 = 50$ N that the kinetic-energy distribution decays at a roughly exponential rate: $\bar{K}_i \propto e^{-0.54 i}$ for 
$i\in \{2,\dots ,10\}$ (see the inset of Figure~\ref{f3} and the Supplementary Information for more details).
Due to this exponential decay, the kinetic energy is reduced by two orders of magnitude over a range of about $10$ sites of the chain. We also observe localization around the second particle. This particular location arises from a combination of dissipation effects and the particular disordered configurations that we examine (a sub-ensemble in which the first particle is always aluminum). We have observed numerically that for the sub-ensemble in which the first particle is tungsten-carbide, the localization phenomenon instead tends to occur around the first particle.
For $i\in \{11, \dots ,15\}$, we observe the excitation of a secondary mode that emerges and decays in an irregular manner as function of time (see Figure~\ref{f3}). For $i\gtrapprox 15$, very low-amplitude waves, which are associated primarily with low-frequency linear modes, reach the right side of the chain and are reflected (see the Supplementary Information for more details).

%%%%%%%%%%%%%%%%%%%%%%%%%%%%%%%
%%%%%%%%%%%%%%%%%%%%%%%%%%%%%%%

\vspace{.5cm}

\begin{figure}
\includegraphics[width=\textwidth]{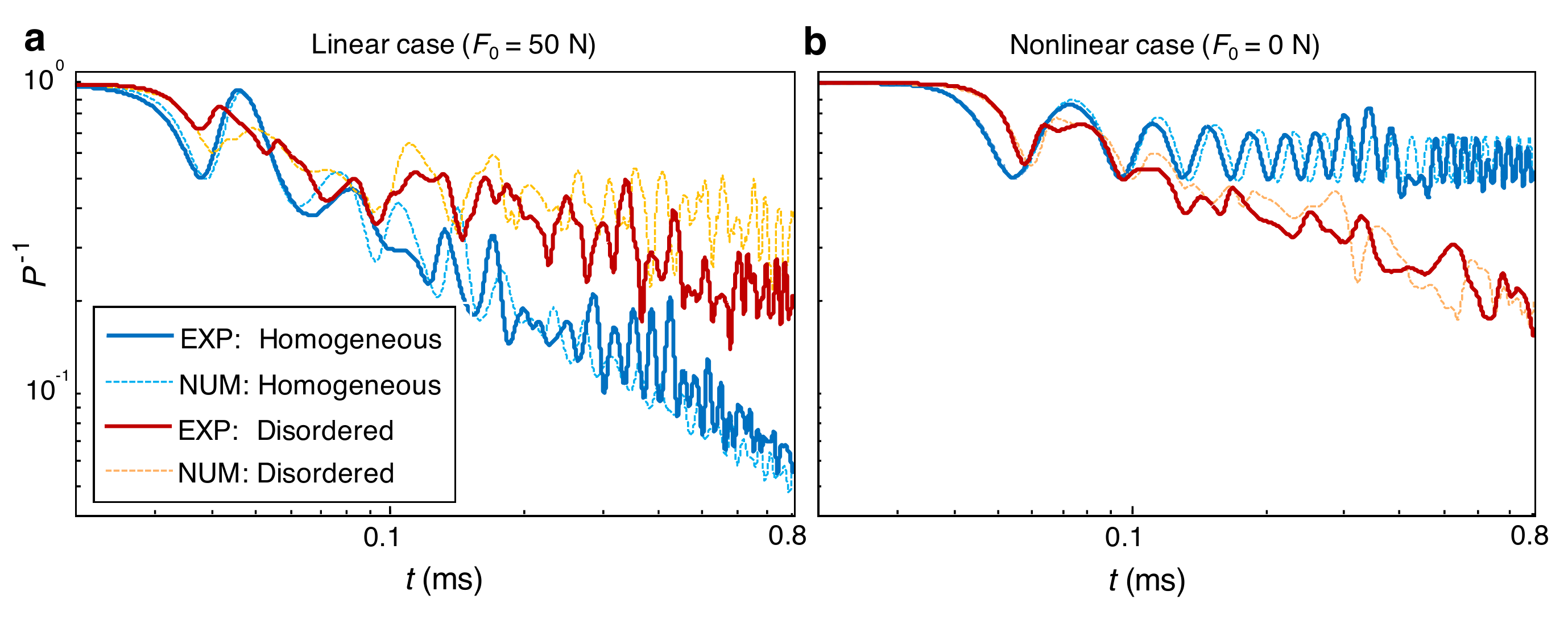}
\caption{\label{f4} {\bf Energy-localization characteristics.} Inverse participation ratio ($P^{-1}$) of the kinetic energy in homogeneous (blue curves) and disordered (red curves) chains for ({\bf a}) linear and ({\bf b}) nonlinear regimes. In both cases, we compare numerical calculations (dashed curves) with experimental data (solid curves).
  }
\end{figure}

\noindent \textbf{Energy localization and spreading amidst nonlinearity and disorder} \\ \\
To investigate the energy transport characteristics of the granular chains, we quantify energy localization and the speed of energy spreading using the inverse participation ratio ($P^{-1}$) and the second moment
of the energy ($m_2$), respectively\cite{ref4,ref5,ref6}. 
To calculate these quantities, we use the kinetic energy instead
of the total energy of the particles in a chain, because we can directly calculate the former experimentally by measuring particle velocities. 
The $P^{-1}$ based on the kinetic energy is
\begin{equation}
 	P^{-1}(t) = \frac{\sum_{i=1}^{N}\left(m_i v_i^2\right)^2}{\left(\sum_{i=1}^{N}m_i v_i^2\right)^2}\,.
\end{equation}
When all energy is confined to a single particle, $P^{-1} = 1$, and $P^{-1}$ approaches $1/N$ (where $N$ is the total number of particles in a chain) as a wave disperses. In Figure~\ref{f4}, we show $P^{-1}$ as a function of time in both the linear and nonlinear regimes. Because of the customary exchange between kinetic and potential energies, we observe oscillations in the temporal 
profile of $P^{-1}$. However, the aggregate trends do not vary significantly from those based on the total energy (see Supplementary Information).

%%%%%%%%%%%%%%%%%%%%%%%%%%%%%%%
%%%%%%%%%%%%%%%%%%%%%%%%%%%%%%%

In a homogeneous chain with precompression (solid blue curve in Figure~\ref{f4}a), we observe experimentally that $P^{-1}$ decreases in time, because the linear waves disperse as they propagate. However, in a disordered chain (solid red curve in Figure~\ref{f4}a), this decreasing trend is less pronounced. This implies that the disorder tends to favor wave localization, confirming the effect of Anderson localization. Our numerical simulations (dashed curves, see Methods) corroborate the experimental results. 
Strikingly, we observe completely different qualitative behavior in the nonlinear regime (see Figure~\ref{f4}b). The nonlinearity favors wave localization in a homogeneous chain, as indicated by the higher value $P^{-1}$. In this case, a localized solitary wave propagates in a highly localized (doubly exponentially decaying\cite{pego1}) manner, and $P^{-1}$ retains a value of $0.6$. A similar trend was noted recently in numerical simulations\cite{ref4,ref11} for bulk dynamics in very long chains (so boundary effects were neglected). In contrast, in our experiments, boundary effects are fundamental both for the generation of the initial excitation and for the ensuing dynamics.

%%%%%%%%%%%%%%%%%%%%%%%%%%%%%%%
\begin{figure}
\includegraphics[width=\textwidth]{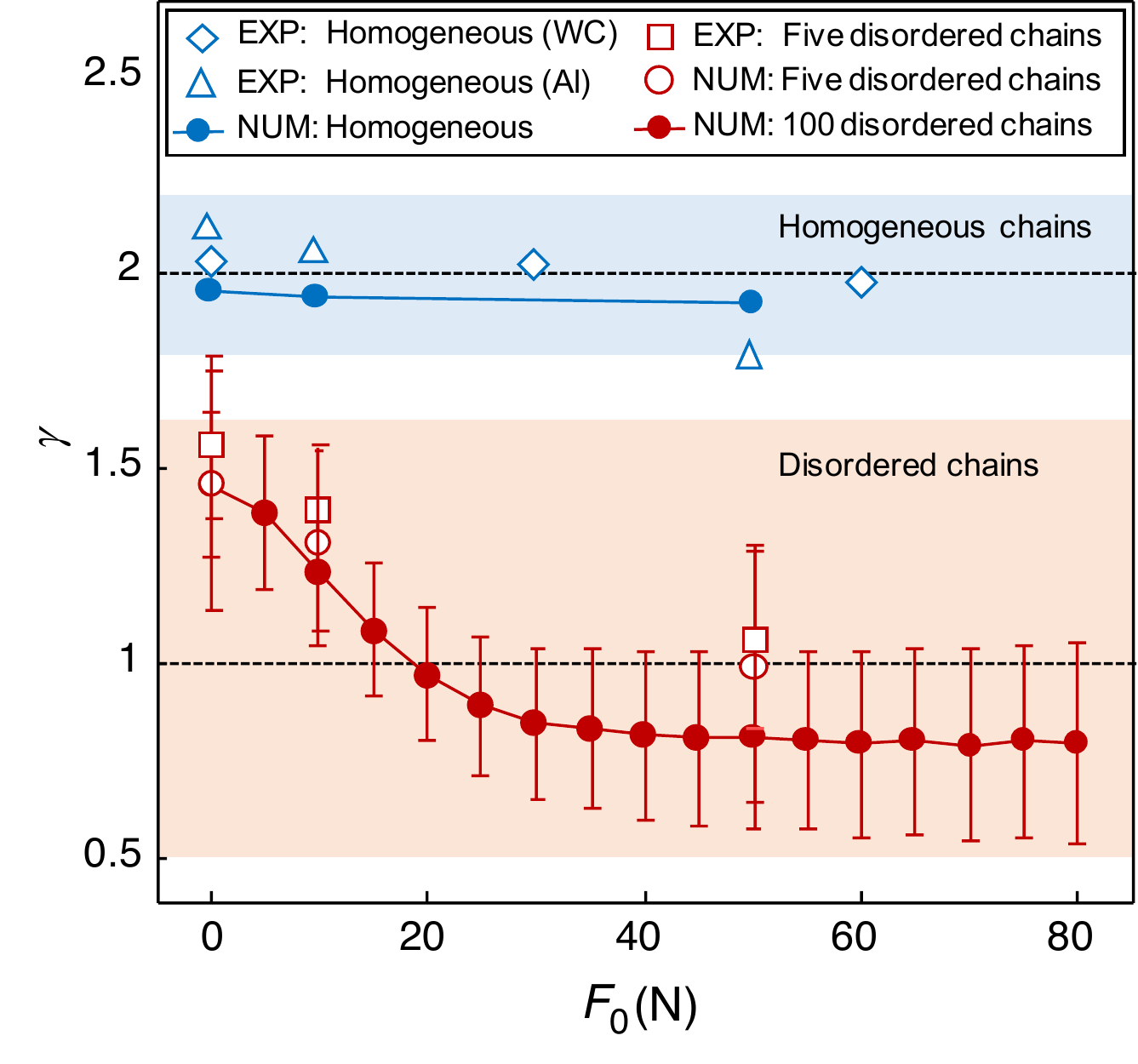}
\caption{\label{f5} {\bf Energy-spreading characteristics.} Exponents ($\gamma$) of the second moment $m_2$ of kinetic energy as a function of precompression strength. The exponents $1$ and $2$ (horizontal dashed lines) represent diffusive and ballistic transport, respectively. The diamond, triangular, and square marks are based on experimental data obtained, respectively, from a homogeneous tungsten-carbide (WC) chain, a homogeneous aluminum (Al) chain, and a mean over five disordered chains. (See the Supplemental Information for details about the five disordered chains.) We mark the numerical data with circles. We do computations for homogeneous chains (blue circles), the five disordered chains (hollow red circles), and 100 randomly generated disordered chains (red circles with error bars). 
}
\end{figure}
%%%%%%%%%%%%%%%%%%%%%%%%%%%%%%%

We obtain the second moment of kinetic energy distribution by calculating
\begin{equation}
 	m_2(t) = \frac{\sum_{i=1}^{N}i^2\left(m_i v_i^2\right)}{\sum_{i=1}^{N} m_i v_i^2}\,.
\end{equation}
It is well-known that asymptotic energy spreading in a homogeneous chain is ballistic (i.e., $m_2(t)\sim t^2$ as $t\rightarrow \infty$)\cite{ref4,ref12}. To investigate energy spreading in detail, we estimate the exponent $\gamma$ of the second moment in the scaling relationship $m_2(t)\sim t^{\gamma}$ during the time period from $0.1$ ms to $1$ ms\cite{ref5,ref4,ref11}. 
We compare the estimated exponents from the experimental data to those from numerical simulations in Figure~\ref{f5}. 
The exponents in homogeneous chains (the upper shaded area in Figure~\ref{f5}a) are about $\gamma=2$ (corresponding to ballistic
spreading). For larger precompression, the exponents are slightly smaller, but they remain near the ballistic regime. 

For disordered chains, we observe drastic changes in $\gamma$ as we increase $F_0$.
Specifically, the mean values of the exponents for the disordered chains diminish gradually from superdiffusive ($1<\gamma<2$) to subdiffusive ($0<\gamma<1$) for chains with progressively stronger static precompressions.
It is important, however, to contrast even the subdiffusive chains, for which we observe $\gamma \in (0.5,1)$, with the much slower
subdiffusive spreading in discrete nonlinear Schr{\"o}dinger
and Klein--Gordon chains\cite{ref5,flach08,flach09}.
Our experimental results (red squares) corroborate our numerical computations (red circles). 

We also compare the exponents that we obtain from the five disordered chains with those from numerical simulations of 100 different disordered chains in Figure~\ref{f5}. The red circles and error bars, respectively, give the mean values and standard deviations of the computed exponents from the 100 cases. The relevant trends persist despite the larger error bars for stronger
precompressions. The agreement between computations and experiments also continues to hold. Additionally, although our study focuses on the effect of nonlinearity and disorder, the boundary conditions in granular chains can also alter energy-transport mechanisms, as demonstrated previously in other physical systems (see, e.g.,\cite{AndersonSurface,MIMsurface}). See the Supplemental Information for detailed numerical studies of such boundary effects.

%%%%%%%%%%%%%%%%%%%%%%%%%%%%%%%%%%%%%%%%%%%%%%%%%%%%%%%%%%%%
%%
%%                D I S C U S S I O N
%%
%%%%%%%%%%%%%%%%%%%%%%%%%%%%%%%%%%%%%%%%%%%%%%%%%%%%%%%%%%%%

\section*{Discussion}

\noindent In this study, we experimentally and numerically investigated energy transport and localization properties in both ordered and disordered one-dimensional granular crystals in strongly nonlinear, weakly nonlinear, and nearly linear settings.
In our experiments, we observe exponential attenuation of the energy distribution for disordered chains when we apply strong precompression to obtain nearly linear dynamics. This is reminiscent of Anderson localization.
We show that this localized pattern oscillates persistently near the chain boundary. For progressively weaker precompression, the system's effective nonlinearity becomes larger, and there is a gradual progression from a localized pattern to flatter energy distributions. This affects the chain's transport properties, as energy spreads faster for weaker precompression.

Our study paves the way towards using granular crystals and related systems (e.g., ones involving magnets\cite{leonard}) as accessible platforms for exploring the interplay between nonlinearity and disorder in a lattice setting. The amenability of such granular metamaterials for controllable tuning between strongly nonlinear and nearly linear regimes, coupled with the ability to monitor such systems in a distributed fashion, promises a wealth of exciting advances in condensed-matter physics, materials science, and nonlinear dynamics. Proposals to engineer local nonlinearity effects (see, e.g.,\cite{guillaume}) in granular crystals and to examine the competition between the chiefly superdiffusive effects of inter-site nonlinearities and the apparently chiefly subdiffusive effects of on-site nonlinearities also constitute important future directions.

%%%%%%%%%%%%
%\section*{Methods}

%\noindent Methods and associated references are available in the online version of the paper.

%%%%

\bibliographystyle{naturemag}

%{\bf \noindent References\\}

%%%%%%%%%
\section*{Acknowledgements}

\noindent A.~J.~M. acknowledges support from CONICYT (BCH72130485/2013). J.~Y. thanks the NSF (CAREER-1553202) for financial support. J.~Y. and P.~G.~K. also acknowledge support from US-ARO under grant (W911NF-15-1-0604), and P.~G.~K. also gratefully acknowledges support from the Stavros Niarchos Foundation via the Greek Diaspora Fellowship Program. E.K. acknowledges the support from the National Research Foundation of Korea (NRF) grant funded by the Korea government (MSIP, No. 2017R1C1B5018136)

%%%%%%%%%
\section*{Contributions}

All authors wrote the paper and contributed equally to the production of the manuscript and interpretation of results; P.~G.~K., A.~J.~M., M.~A.~P., and J.~Y. designed the study; E.~K. and S.~E.~P. performed the experiments; and E.~K., A.~J.~M., and S.~E.~P. did the numerical computations and data analysis.

%%%%%

\section*{Competing financial interests}

The authors declare no competing financial interests.

%%%%%%

%\bibliographystyle{unsrt}
%\bibliography{Biblio6}

\clearpage
\section*{Methods}

%%%%%%%%%%%%%%%%%%%%%%%%%%%
\header{Experiments} 
We measure wave propagation in two types of one-dimensional granular crystals 
(i.e., granular chains): homogeneous and disordered chains. 
Each chain consists of 32 particles using either tungsten-carbide
(Young modulus $E=600$ GPa, Poisson ratio $\nu=0.2$, and density $\rho=15.6$ 
g/cm$^3$) or aluminum ($E=69$ GPa, $\nu = 0.33$, and $\rho = 2.8$ g/cm$^3$). The 
radius of each particle is $9.53$ mm.
The values of $E$, $\nu$, and $\rho$ are based on standard 
specifications\cite{ref17}. We examine two homogeneous chains (one made of 
tungsten-carbide and the other made of aluminum) and five 
disordered chains (with combinations of tungsten-carbide and aluminum as described 
in Supplementary Note 1). Each particle in a disordered chain except for the 
first particle is randomly chosen between tungsten-carbide and aluminum 
particles, with an independent $50\%$ probability of each for each particle. This yields
an ``uncorrelated'' type of disorder\cite{ref4}. 

The spherical particles are supported by two stainless steel rods coated by 
polytetrafluoroethylene (PTFE) tape to reduce friction between the particles and 
the supporting rods. We also place two aluminum rods on top of the 
granular chain, with minimum clearance to restrict lateral motion of the 
particles. To apply various precompression strengths to a chain, we press a 
heavy block on a sliding guide towards the chain using a linear spring.
Between the block and the spherical chain, we embed a static force sensor 
(see Figure~\ref{f1}) to ensure that the precompression applied to the chain is 
monitored accurately. The other side of the chain is blocked by a steel
plate with a circular hole in the center, and we bond four equidistant steel 
balls to the edge of the hole (see the inset of Figure~\ref{f1}).
 These balls have point contacts with the first particle; this
yields a much smaller contact damping
between the first particle and the plate boundary than would a line 
contact between the two.

For this study, we test granular chains with various precompression strengths (from 
$0$ N to $60$ N).
We give an impact excitation to the first particle using a chrome steel ($E=210$ 
GPa, $\nu=0.29$, and $\rho=7.8$ g/cm$^3$) particle with a $2.38$ mm radius.
 We roll the particle down in a ramp so that it has a normal impact on the first particle
through the hole of the plate (see Figure~\ref{f1}). The impact velocity measured 
by a laser Doppler vibrometer (Polytec OFV 534) %\cite{ref13} 
is $0.45\pm 0.12$ m/s.

To visualize wave propagation in a granular chain, we measure the axial 
velocities of particles individually for each particle spot using the LDV. To synchronize the measured data, we use a small 
piezoelectric ceramic plate (3 mm $\times$ 4 mm $\times$ 0.5 mm) bonded to the first particle; this plate generates a trigger 
signal (voltage) at the instant of striker impact.  We scan the velocity 
profiles for all particles in the chain by moving the 
LDV using a moving stage, and we reconstruct the measured data to depict a 
spatiotemporal profile of propagating waves. For each configuration, we follow the above procedures and conduct 
three tests. This amounts to 672 experimental realizations to construct the 
spatiotemporal dynamics of the chain for all of the cases studied in the present 
article.

\vspace{.5cm}

%%%%%%%%%%%%%%%%%%%%%%%%
\header{Numerical simulations} 
The equation of motion for the $i$th particle in a granular chain is\cite{sen08}
\begin{align}
 		m_i\frac{d^2 u_i}{dt^2} &= 
A_{i-1,i}\left[\delta_{i-1,i}+u_{i-1}-u_{i}\right]_{+}^{3/2} \nonumber\\
 	&- A_{i,i+1}\left[\delta_{i,i+1}+u_{i}-u_{i+1}\right]_{+}^{3/2} - 
\frac{m_i}{\tau}\frac{du_i}{dt}\,,
\end{align}
where $m_i$ and $u_i$ are mass and displacement of the $i$th particle. The force 
on the $i$th particle depends on the geometry and material properties of it and 
its adjacent particles. The interaction coefficient between the $i$th and 
$(i+1)$th particles is
$A_{i,i+1} = 
\frac{4E_{i}E_{i+1}}{3\left[E_{n}(1-\nu_{n-1}^2)+E_{n-1}(1-\nu_{n}^2)\right]}
\sqrt{\frac{R_{i}R_{i+1}}{R_{i}+R_{i+1}}}$, where $E_i$, $\nu_i$, and $R_i$ are, 
respectively, the $i$th particle's Young modulus, Poisson ratio, and radius. The 
quantity $\delta_{i,i+1}=\left(F_0/A_{i,i+1}\right)^{2/3}$ is the compression 
distance between the $i$th and $(i+1)$th particles at static equilibrium under 
static precompression $F_0$. The bracket $\left[x\right]_{+} = 
{\text{max}}\{0,x\}$ encodes the fact that there are no tensile forces in 
interparticle interactions. The damping coefficient $1/\tau$ is determined by 
measuring the decay of leading waves in experiments and determining a parameter 
that matches the experimental results (see the Supplemental Information for more 
details). For the homogeneous chain of aluminum particles, we use $\tau = 2$ ms (for $F_0 = 0$ N), $\tau = 1$ ms (for $F_0 = 10$ N), and $\tau = 0.4$ ms (for $F_0 = 50$ N).
  For the disordered chains, we use $\tau = 1$ ms (for $F_0 = 0$ N), $\tau = 
0.6$ ms (for $F_0 = 10$ N), and $\tau = 0.4$ ms (for $F_0 = 50$ N). 
The equation of motion for the striker particle ($i=0$) includes an interparticle 
interaction only with the first particle, and 
the first particle has an interaction term with the boundary plate and the 
second particle. Similarly, the last particle has an interaction with the penultimate particle 
and the other boundary plate.
The displacements at the boundaries are fixed ($u_{\text{left}}= 
u_{\text{right}}= 0$), and we apply an initial velocity to the striker 
($v_{\text{striker}} = 0.45$ m/s). We solve these equations of motion 
numerically with a Runge--Kutta method using {\sc Matlab}'s {\sc ODE45} routine 
with an relative error tolerance of 0.1\% and absolute error tolerances of $1 \times 10^{-6}$ m for displacement and $1 \times 10^{-6}$ m/s for velocity.

%%%%%%%%%%%%%%%%%%%%%

\renewcommand{\thepage}{S\arabic{page}}  
\renewcommand{\thesection}{S\arabic{section}}   
\renewcommand{\thetable}{S\arabic{table}}   
\renewcommand{\thefigure}{S\arabic{figure}}

%%%%%%%%%

\section*{Supplementary Note 1: {\bf Numerical and experimental results from homogeneous and disordered granular chains, and the detailed configurations of the disordered chains.}}

\noindent In Figure~\ref{f2}, we show the full-field particle velocities for a homogeneous chain that consists of aluminum particles only (top two rows) and a disordered chain composed of aluminum and tungsten-carbide particles (bottom two rows) in strongly nonlinear, weakly nonlinear, and nearly linear scenarios (in the left, middle, and right panels, respectively). In each group, the upper and lower rows show our experimental and numerical results, respectively. Our numerical results corroborate our experimental measurements. 

In Figure~\ref{f2} and also in Figure 2 of the main manuscript, the disordered-chain results are for the first configuration (``Chain 1'') of disordered chains from Table \ref{tab1}. 
We assemble each disordered chain by randomly choosing each individual particle (except for the first one) in the chain as either aluminum or tungsten-carbide with equal probability.
This splits the statistical ensemble into two sub-ensembles: (i) chains with the first particle fixed as aluminum and (ii) chains with the first particle fixed as tungsten-carbide. 
We intentionally choose to examine case (i) because it facilitates transfer of mechanical energy from the striker particle to the chain. This observation is based on preliminary experimental observations using both sub-ensembles. In this case, we observe in our experiments mean energy transfer rates from the striker to the chain of 75.5$\%$, 76.1$\%$, and 85.8$\%$ for $0$ N, $10$ N, and $50$ N precompression, respectively. Although we focus our experimental efforts on studying case (i), we show numerically that case (ii) has similar transport properties as case (i). For instance, in both cases,
we observe a transition from subdiffusive to superdiffusive behavior, although the precise values for the energy spreading-rate exponents $\gamma$ differ (See Supplementary Note 9 for more details).

%%%%%%%%%%%%%%%%%%%%%%%%%%%%%%%%%%%%%%%%%%%%%%%%%%%%%%%%%%%%
\begin{table*}
\caption{\label{tab1}%
Configurations of the five disordered chains}
\begin{tabular}{|p{1.9cm}|c|}
\hline
\textrm{Disordered chain}&\textrm{Order of particles (1:Tungsten-Carbide; 2: Aluminum)}\\\hline
Chain 1& [2, 1, 2, 2, 1, 1, 2, 2, 1, 1, 1, 1, 1, 2, 2, 1, 2, 1, 2, 2, 1, 1, 2, 2, 2, 1, 2, 1, 1, 1, 2, 1]\\\hline
Chain 2& [2, 2, 2, 1, 2, 1, 1, 1, 1, 2, 1, 2, 1, 1, 2, 2, 2, 1, 1, 2, 1, 2, 2, 1, 2, 2, 1, 1, 2, 1, 1, 1]\\\hline
Chain 3& [2, 2, 1, 1, 2, 2, 1, 1, 2, 2, 2, 2, 2, 2, 2, 1, 2, 1, 1, 1, 1, 2, 1, 2, 2, 1, 1, 2, 1, 1, 1, 2]\\\hline
Chain 4& [2, 1, 2, 2, 2, 2, 1, 2, 1, 2, 2, 1, 2, 1, 2, 1, 1, 1, 1, 1, 2, 2, 1, 1, 2, 2, 1, 1, 1, 1, 1, 1]\\\hline
Chain 5& [2, 2, 2, 2, 1, 2, 1, 1, 2, 1, 2, 2, 1, 2, 2, 1, 1, 1, 1, 2, 1, 2, 2, 2, 1, 1, 1, 1, 2, 1, 2, 1]\\\hline
\end{tabular}
\end{table*}
%%%%%%%%%%%%%%%%%%%%%%%%%%%%%%%%%%%%%%%%%%%%%%%%%%%%%%%%%%%%

\begin{figure}
\includegraphics[width=\textwidth]{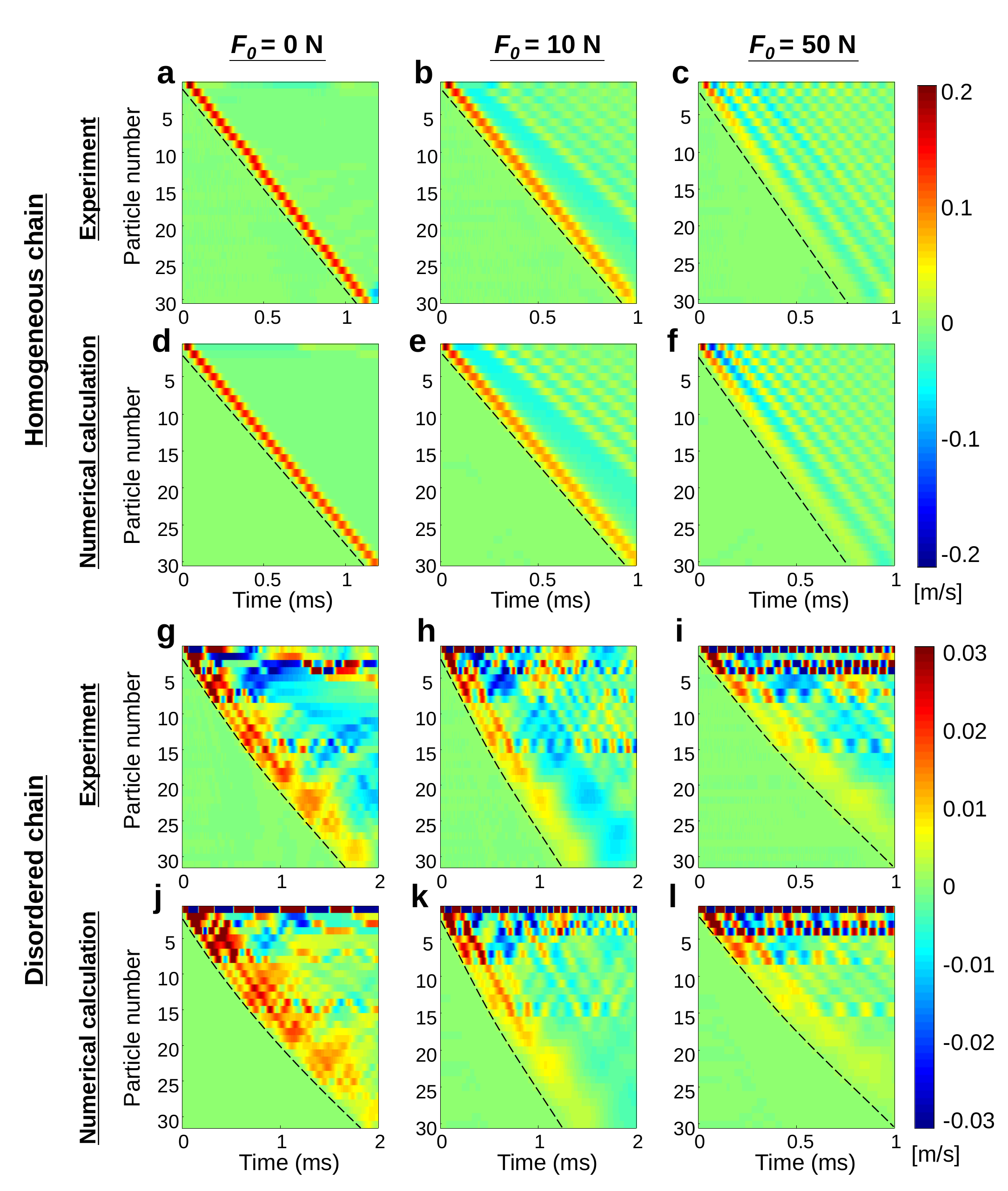}
\caption{\label{f2} {\bf Wave propagation in homogeneous and disordered chains.} ({\bf a--f}) Spatiotemporal distributions of particle velocities in a homogeneous chain consisting of aluminum particles with static precompressions
of (a,d) $0$ N, (b,e) $10$ N, and (c,f) $50$ N. We compare the experimental data (panels {\bf a--c}) to numerical calculations (panels {\bf d--f}). ({\bf g--l}) Spatiotemporal distributions of particle velocities in a disordered chain (the ``Chain 1'' configuration
in Table \ref{tab1}) with static precompressions of (g,j) $0$ N, (h,k) $10$ N, and (i,l) $50$ N. We compare the experimental data (panels {\bf g--i}) to numerical calculations (panels {\bf j--l}). We show dashed arcs to visually track the edges of the velocity distributions.
}
\end{figure}

%%%%%%%%%%%%%%%%%%%%%%%%%%%%%%%
\section*{Supplementary Note 2: {\bf Nonlinearity of stress waves propagating in disordered chains.}}

\begin{figure}
\includegraphics[width=\textwidth]{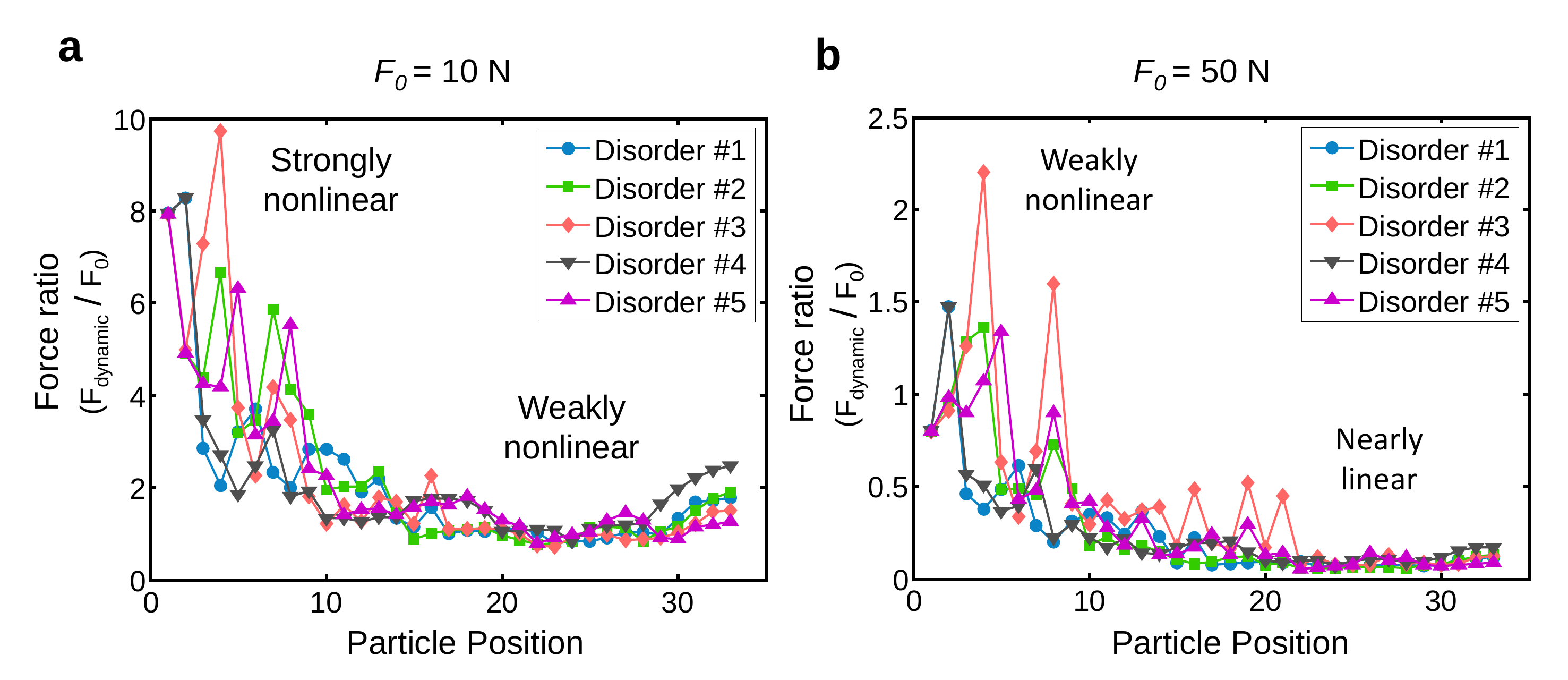}
\caption{\label{fs7} {\bf Ratio of maximum dynamic force to static precompression.} ({\bf a}) Ratio of the maximum dynamic force (defined as $F_{\text{dynamic}}=\max(|F-F_0|)$, where $F$ is the propagating wave's force amplitude)
  to the static precompression $F_0$ in disordered chains for $F_0 = 10$ N . ({\bf b}) Ratio of maximum dynamic force to
the static precompression in disordered chains for $F_0 = 50$ N.}
\end{figure}

\noindent In a disordered granular chain, stress waves scatter as they propagate along the chain, which in turn results in a diminution of the wave amplitude. When precompression is applied to a chain, this amplitude decrease
changes the nonlinearity strength in the system as the stress waves propagate along the chain. One can estimate the nonlinearity strength of the propagating wave based on the ratio of its dynamic force to the static precompression. We use the following qualitative
criterion for characterizing regimes with different nonlinearity strengths. We say that dynamics is ``strongly nonlinear'' when the force ratio $F_{\text{dynamic}}/F_0 \gg 1$, ``weakly nonlinear'' when $F_{\text{dynamic}}/F_0 \approx 1$, and ``nearly linear'' (or sometimes just ``linear'' as a shorthand) when $F_{\text{dynamic}}/F_0 \ll 1$. The maximum dynamic force is defined as $F_{\text{dynamic}}=\max(|F-F_0|)$, where $F$ is the propagating wave's force amplitude and $F_0$ is the static precompression. 
In Figure~\ref{fs7}a, we show $F_{\text{dynamic}}/F_0$ when $F_0=10$ N.
Initially, a strongly nonlinear wave is formed, but it becomes a weakly nonlinear one due to scattering and localization.
 In Figure~\ref{fs7}b, we show the force ratio for $F_0 = 50$ N precompression, where
a weakly nonlinear wave becomes a nearly linear one as it propagates.

%%%%%%%%%%%%%%%%%%%%%%%%%%%%%%%
%%%%%%%%%%%%%%%%%%%%%%%%%%%%%%%

\section*{Supplementary Note 3: {\bf Linear modes of a disordered chain and its frequency response function.}}

\begin{figure}
\includegraphics[width=\textwidth]{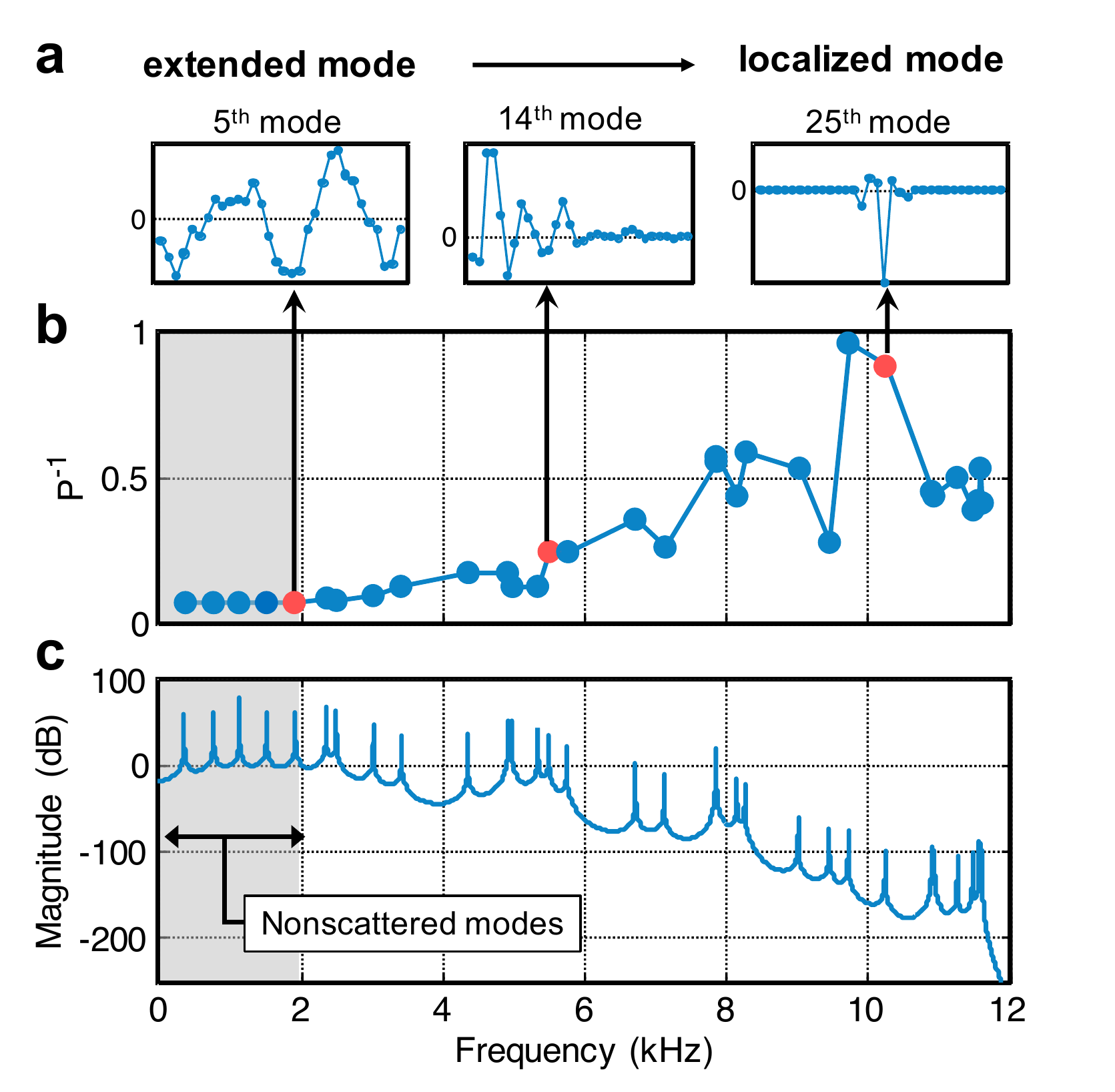}
\caption{\label{fs9} {\bf Vibration modes of a disordered chain (Chain 1 from Table 1) for a static precompression of $F_0 = 50$ N.} ({\bf a}) Mode profiles of the $5^{\text{th}}$, $14^{\text{th}}$, and $25^{\text{th}}$ modes. ({\bf b}) Inverse
participation ratio (IPR) $P^{-1}$ calculated based on the displacement amplitude for each mode. ({\bf c}) Frequency response function of the disordered chain. The shaded area shows the nonscattered modes
(from the $1^{\text{st}}$ to the $5^{\text{th}}$ modes).}
\end{figure}

\noindent When precompression is strong enough compared to the dynamic force,
one can linearize Eq.~(3) in the main text. Using this linearization, one gives an approximate description of a granular chain's dynamics as a superposition of the temporal evolution of different vibration modes. The weight of each mode depends on the initial excitation. For a disordered chain, vibration modes with predominantly low frequencies are delocalized, producing what are sometimes called ``nonscattered modes''\cite{ref3}. The number $q_{\text{ns}}$ of nonscattered modes satisfies $q_{\text{ns}}\sim \sqrt{N}$ as $N\rightarrow \infty$, where $N$ is the total number of particles in the chain. In our 32-particle chains, roughly the first six modes are nonscattered modes. Consequently, the waves with frequencies associated with nonscattered modes tend to propagate along the chain without localization. For progressively larger frequencies, vibration modes become localized, reducing the effective number of particles that oscillate with an amplitude that differs substantially from $0$. We quantify this phenomenon by computing the inverse participation ratio (IPR) $P^{-1}$, which we show in Figure~\ref{fs9}. 
We also show representative vibration modes and frequency response functions of a disordered chain (using Chain 1 from Table 1) at $F_0=50$ N. Because the IPR represents the degree of localization, 
an extended mode --- corresponding
to a low-frequency wave mode with a large wavelength --- has a low $P^{-1}$ value, whereas a localized mode has $P^{-1}$ close to $1$.

%%%%%%%%%%%%%%%%%%%%%%%%%%%%%%%
%%%%%%%%%%%%%%%%%%%%%%%%%%%%%%%

%%%%%%%%%%%%
\section*{Supplementary Note 4: {\bf Exponential fit of kinetic-energy decay in the Anderson-like mode.}}

To support the exponential fit that we used to characterize the decay of the kinetic-energy distribution (see Figure~3 in the main manuscript), we perform an Anderson--Darling test\cite{ADtest}. 
Because we estimate the parameter of the exponential trend from data, to 
implement the test, we apply a Monte-Carlo procedure with 1000 data sets generated 
under the null hypothesis of the exponential distribution $\bar{K}_i \propto e^{-0.54 i}$. Using particles 
$i\in \{2,\dots ,10\}$, we obtain a p-value of 0.96, so it passes the standard statistical test indicating that there is no significant
departure from normality.

%%%%%%%%%%%%
\section*{Supplementary Note 5: {\bf Comparison of $\gamma$ between kinetic energy and total energy.}}

\begin{figure}
\includegraphics[width=\textwidth]{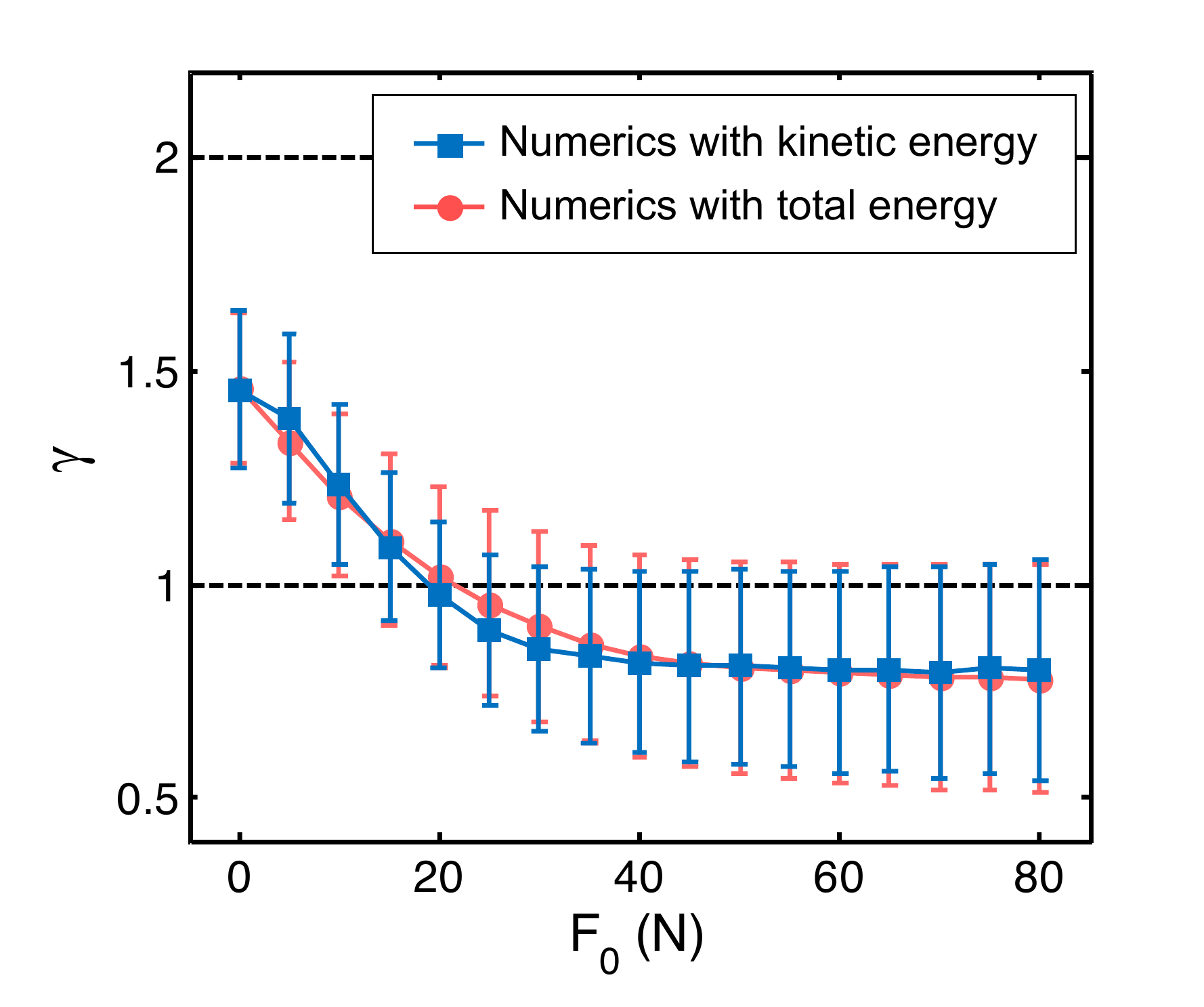}
\caption{\label{fs2} {\bf Comparison of second moments from kinetic energy
    versus those from total energy for different precompression strengths.} The blue squares represent calculations based on kinetic energy, and the red circles represent calculations based on total energy (i.e., kinetic energy
plus potential energy). The vertical axis gives the exponent $\gamma$ of the second moment, and the horizontal axis gives the static precompression strength.}
\end{figure}

\noindent In Figure~\ref{fs2}, we compare the exponents $\gamma$ of the second moment ($m_2(t)\sim t^\gamma$) of the
kinetic energy with those that we calculate using the total
energy. This figure illustrates that, despite the
observed oscillations of $m_2(t)$
because of the kinetic--potential energy exchange in the propagating waves, the exponents of the experimentally
measurable kinetic energy are essentially the same as those
obtained from total energy.

%%%%%%%%%%%%%%%%%%%%%%%%%%%%%%%
%%%%%%%%%%%%%%%%%%%%%%%%%%%%%%%

\section*{Supplementary Note 6: {\bf Types of initial conditions: velocity and displacement excitations}}

We consider two types of perturbations. Similar to what was done previously in\cite{ref1,ref2} for the
bulk of a granular chain, in our examination of boundary-induced excitations,
we conduct simulations with two specific initial conditions.
We apply a {\it velocity} excitation, which consists of an initial perturbation of the velocity of a single particle, with all other particles starting with zero velocity and all particles starting with zero displacements (see Figure~\ref{fs4}a).
We also consider a {\it displacement} excitation, which is a perturbation of the position from equilibrium of a single particle (see Figure~\ref{fs4}b). Further discussions follow next along with the investigations on the short and long chain configurations.

%%%%%%

\section*{Supplementary Note 7: {\bf Comparison of $m_2$ between short and long chains.}}

\begin{figure}
\includegraphics[width=\textwidth]{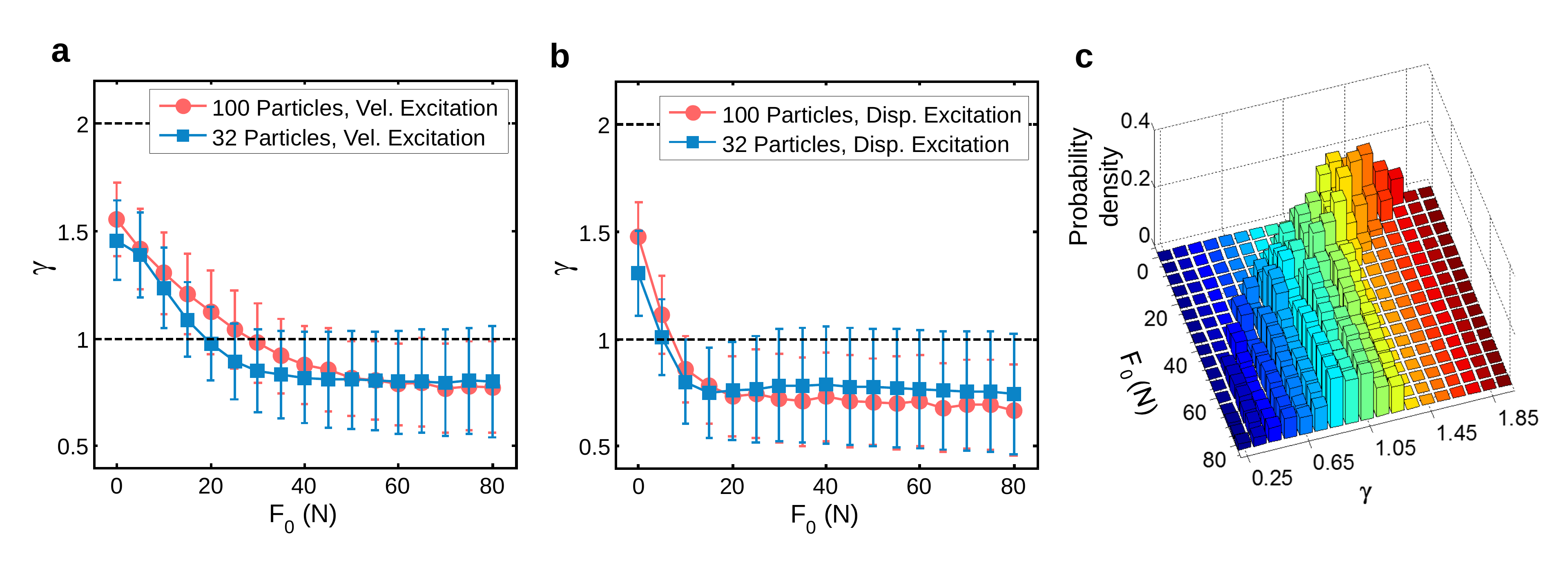}
\caption{\label{fs4} {\bf Comparison of exponents $\gamma$ of the second moment $m_2$ of the kinetic energy in short versus long chains.} We compare the exponents between short chains (32 particles) and long chains (100 particles) for ({\bf a}) a velocity excitation and ({\bf b}) a displacement excitation. We use 100 chains for each case, and we generate them randomly by independently choosing each particle in the chain as either tungsten-carbide and aluminum particles with equal probability. ({\bf c}) Probability density as a function of statistic precompression $F_0$ and spreading exponent $\gamma$.
}
\end{figure}

\noindent We use numerical simulations to compare the exponent $\gamma$ of the second moment $m_2$ of the kinetic energy between short (32 particles, square) and long (100 particles, circles) chains. We do this comparison with both types of velocity and displacement excitations. Previous studies reported that $\gamma$ achieves a specific value\cite{ref1,ref2,ref3} faster in time with
velocity excitations than with displacement excitations\cite{ref1,ref2}. In our simulations, we observe that the variation of the exponents is relatively small after about $0.1$ ms, similar to the trend reported in ~\onlinecite{ref1}. Therefore, we estimate the exponents during the time periods $[0.1,1]$ ms for the short chains and during $[0.1,3]$ ms for the long chains using a least-squares fit. As we show in Figures~\ref{fs4}a (velocity excitations) and ~\ref{fs4}b (displacement excitations), the results from the short chains have minor discrepancies from those for the long chains, especially when one considers the large standard deviation (see the error bars) of the extracted exponents. For reference, Figure~\ref{fs4}c shows the statistical distribution of the exponents calculated using $100$ chains of $32$ particles for a velocity excitation (corresponding to the curve with blue squares in Figure~\ref{fs4}a). For weak precompression, the long chains have relatively large exponents for both velocity and displacement excitations. We expect that this observation is related to the detrapping of localized energy in the weakly nonlinear regime\cite{ref2}. We also observe that the exponents for the displacement excitations are smaller than corresponding ones for the velocity excitation (compare Figs.~\ref{fs4}a and \ref{fs4}b). This arises because a displacement excitation contains broadly-distributed frequencies, whereas a velocity excitation contains mostly low-frequency signals. In Supplementary Note 8, we further discuss boundary effects.

%%%%%%%%%%%%%%%%%%%%%%%%%%%%%%%

%%%%%%%%%%%%%%%%%%%%%%%%%%%%%%%

\section*{Supplementary Note 8: {\bf Boundary effects}}

\begin{figure}
\includegraphics[width=\textwidth]{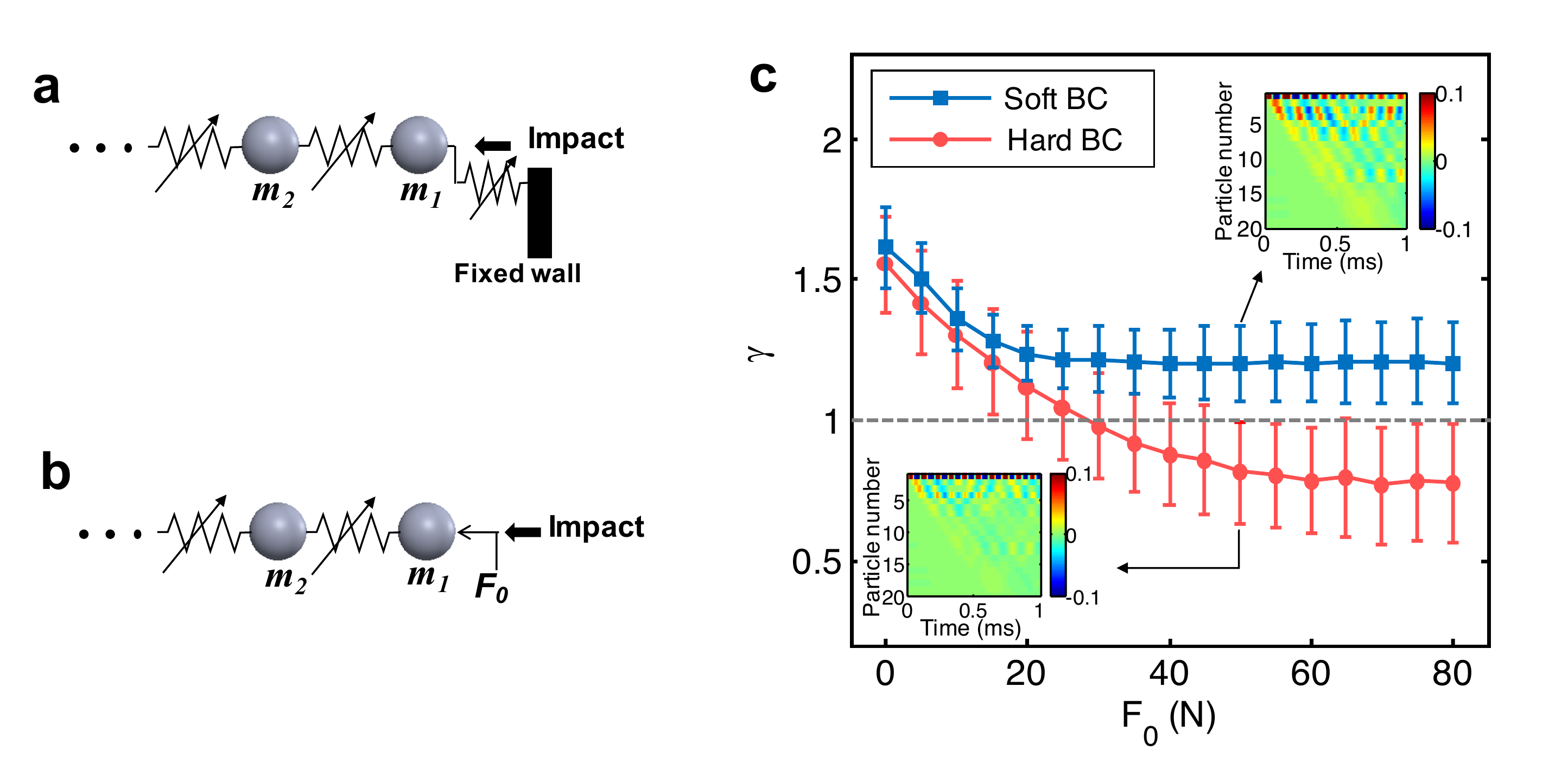}
\caption{\label{fs3} {\bf Boundary conditions at the location of the excitation and their corresponding effects on energy transport.} ({\bf a}) Schematic of an excitation as a hard boundary constrained by a fixed wall. ({\bf b}) Schematic of an excitation as a
  soft boundary (in particular, a free boundary) in which precompression
is applied directly to the first particle without any constraints. In both panels, the depicted springs represent
Hertzian interactions. ({\bf c}) Comparison
of the second moment $m_2$ of the kinetic energy, averaged over 100 realizations, for granular chains with two different boundary conditions at the right end: a hard boundary (red circle) and soft boundary (blue square). The insets represent spatiotemporal distributions of propagation velocity for particles in chains with the two different 
boundary conditions.
}
\end{figure}

Previous studies reported that when the disorder is uncorrelated (as in our case),
velocity excitations in the bulk exhibit superdiffusive transport regardless of the nonlinear regime,
whereas displacement excitations exhibit a transition from subdiffusive transport in the linear regime to superdiffusive spreading in the strongly nonlinear regime\cite{ref1,ref2}. 
Interestingly, when we excite the first particle at the fixed boundary of a granular chain, we observe that both velocity and displacement excitations exhibit a transition from subdiffusive to superdiffusive transport
(see Figure~\ref{fs4}). To investigate the effect of boundary conditions on energy-transport mechanisms, we conduct numerical simulations using 100-particle chains and compare two scenarios. One scenario is a hard boundary fixed with a massive
wall (as in the experimental setup), and the other is a soft
boundary in which we apply a consistent precompression force to the
first particle (see Figures~\ref{fs3}a,b). For a fixed boundary with a wall, the boundary particle ($i=1$) interacts with three objects: the striker ($i=0$), an adjacent particle ($i=2$), and the wall
($i=w$). Its equation of motion is
\begin{align}\label{one}
 	m_1\frac{d^2 u_1}{dt^2} &= A_{0,1}\left[\delta_{0,1}+u_{0}-u_{1}\right]_{+}^{3/2}- A_{1,2}\left[\delta_{1,2}+u_{1}-u_{2}\right]_{+}^{3/2}\nonumber\\
 	&\quad+A_{w,1}\left[\delta_{w,1}-u_{1}\right]_{+}^{3/2}- \frac{m_1}{\tau}\frac{du_1}{dt}\,,
\end{align}
where first and second terms on the right-hand side represent the interactions of the first particle with the striker and the second particle, respectively. The third term on the right-hand side models the interaction with the fixed wall, and the last term represents dissipation 
in the first particle.
For a free boundary with a constant force, the third term in Eq.~\eqref{one}, which represents the interaction with the wall, is replaced with a constant force $F_0$. This yields the equation
\begin{align}
	 m_1\frac{d^2 u_1}{dt^2} &= A_{0,1}\left[\delta_{0,1}+u_{0}-u_{1}\right]_{+}^{3/2}- A_{1,2}\left[\delta_{1,2}+u_{1}-u_{2}\right]_{+}^{3/2}\nonumber\\
	 &\quad+F_0- \frac{m_1}{\tau}\frac{du_1}{dt}\,.
\end{align}
One can construe such a free boundary as an extreme case of a soft boundary. 

We conduct numerical simulations for both boundary conditions. In Figure~\ref{fs3}c, we observe that the choice of boundary condition has a negligible effect for weak precompression, and the energy spreading exhibits similar superdiffusive behavior for both hard and soft boundaries. However, for progressively stronger precompression, the disparity between the two scenarios becomes larger.
For hard boundaries, a significant amount of excitation energy is
trapped near the boundary for a long time
(compare the insets of Figure~\ref{fs3}c),
so the exponent $\gamma$ of the second moment suggests a trend towards subdiffusive spreading as one increases precompression. 
In contrast, for a soft boundary, localization is weaker and the
energy spreads more rapidly. The resulting mean exponent $\gamma$ of $m_2$
asymptotically approaches a particular value (roughly $1.2$) in the superdiffusive regime. This asymptotic value depends on the
mass ratio of the heavy and light particles, similar to what has been described for the bulk of a granular chain\cite{ref2}, and also on the
initial excitation. Note that we estimate values of $\gamma$ using a
specified finite duration in both our numerical computations and our experiments, so one can expect some discrepancies between these values and ones that are calculated over an infinite time 
horizon (i.e., asymptotic values as $t\rightarrow\infty$). However, in past work\cite{ref1}, it has been observed in numerical computations that the value of $\gamma$ in an initial time period 
(of about $0.1$--$1$ ms) seems to persist for a long time without significant changes.

Lastly, comparing the results from Figure~\ref{fs3}c with those from Figure 5 of the main manuscript, we can deduce that the boundary condition of our experimental setup corresponds to the hard boundary condition (i.e., the setup shown in Figure~\ref{fs3}a), based on the fact that $\gamma \leq 1$. This makes sense, given the clamped configuration of the first particle in the chain as shown in the inset of Figure 1 of the main manuscript.

%%%%%%%%%%%%%%%%%%%%%%%%%%%%%%%
\section*{Supplementary Note 9: {\bf Effect of the first particle in the chain on the energy spreading.}}

\begin{figure}
\includegraphics[width=\textwidth]{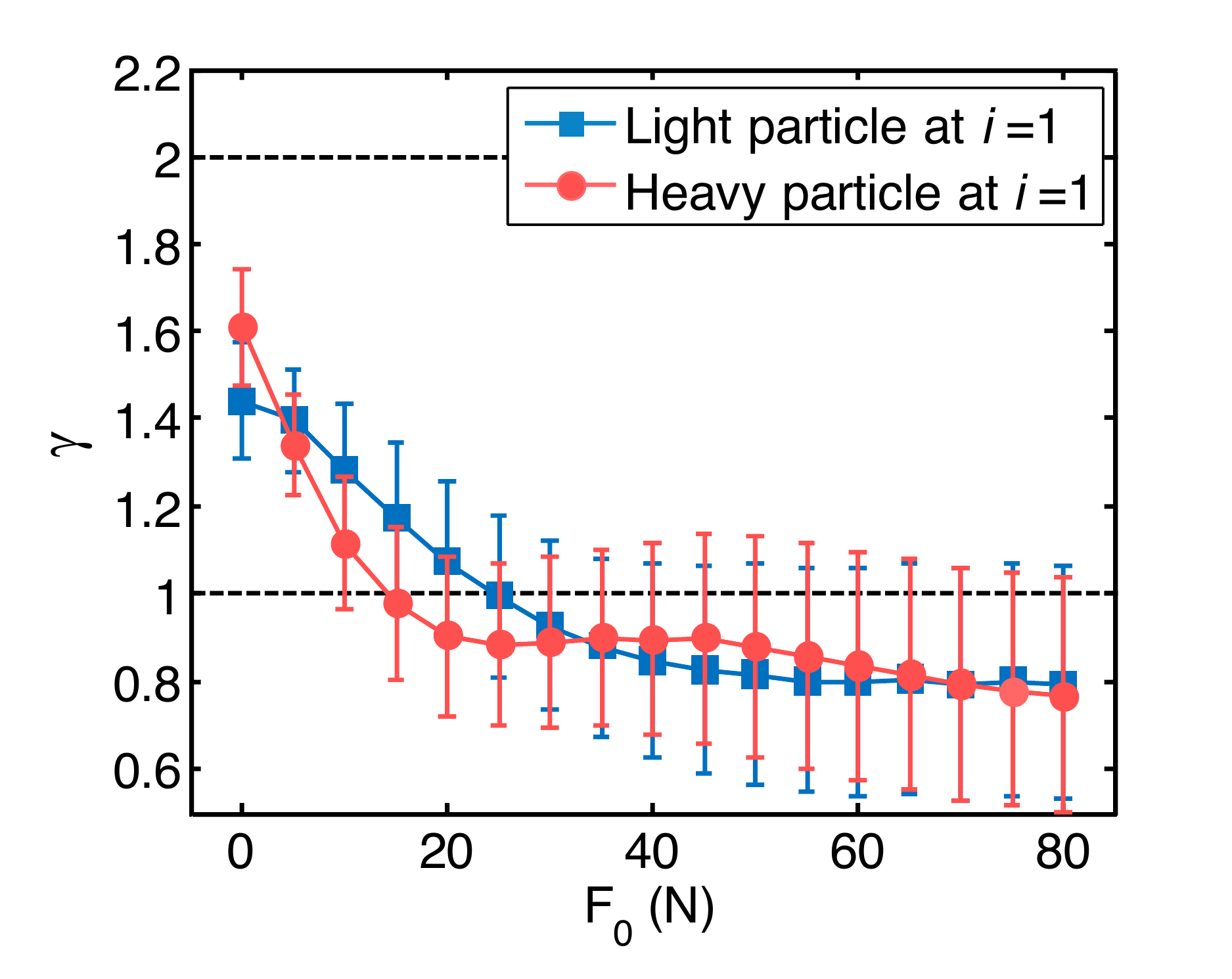}
\caption{\label{fs6} {\bf Effect of the choice of material of the first particle on energy spreading.} We compare the exponent $\gamma$ of the second moment $m_2$ of the kinetic energy between chains
  that have a light particle (aluminum) at $i=1$ (i.e., in the first position) and chains that
  have a heavy particle (tungsten-carbide) at $i=1$.
  All chains have 100 particles and a fixed-wall
  boundary at the right.
 }
\end{figure}

\noindent We compare the exponents $\gamma$ of the second moment $m_2$ of the kinetic energy for two kinds of chains: one group with a light particle (aluminum particle) and the other group with a heavy particle (tungsten-carbide) at the first ($i=1$) spot. In Figure~\ref{fs6}, we show computational results for these two cases from numerical simulations of 100 chains with 100 particles each. For weak precompression, the two chains exhibit very different behaviors. As we consider stronger precompression, however, the exponents of the two groups approach the same value, and the energy in both groups is transported in a subdiffusive way. In the main text, we used light (i.e., aluminum) particles at the $i=1$ spot to facilitate the transport of more energy to the granular chain from the striker particle.

%%%%%%%%%%%%%%%%%%%%%%%%%%%%%%%
%%%%%%%%%%%%%%%%%%%%%%%%%%%%%%%

%%%%%%%%%%%%%%%%%%%%%%%%%%%%%%%
\section*{Supplementary Note 10: {\bf Effect of dissipation on energy spreading.}}

\begin{figure}
\includegraphics[width=\textwidth]{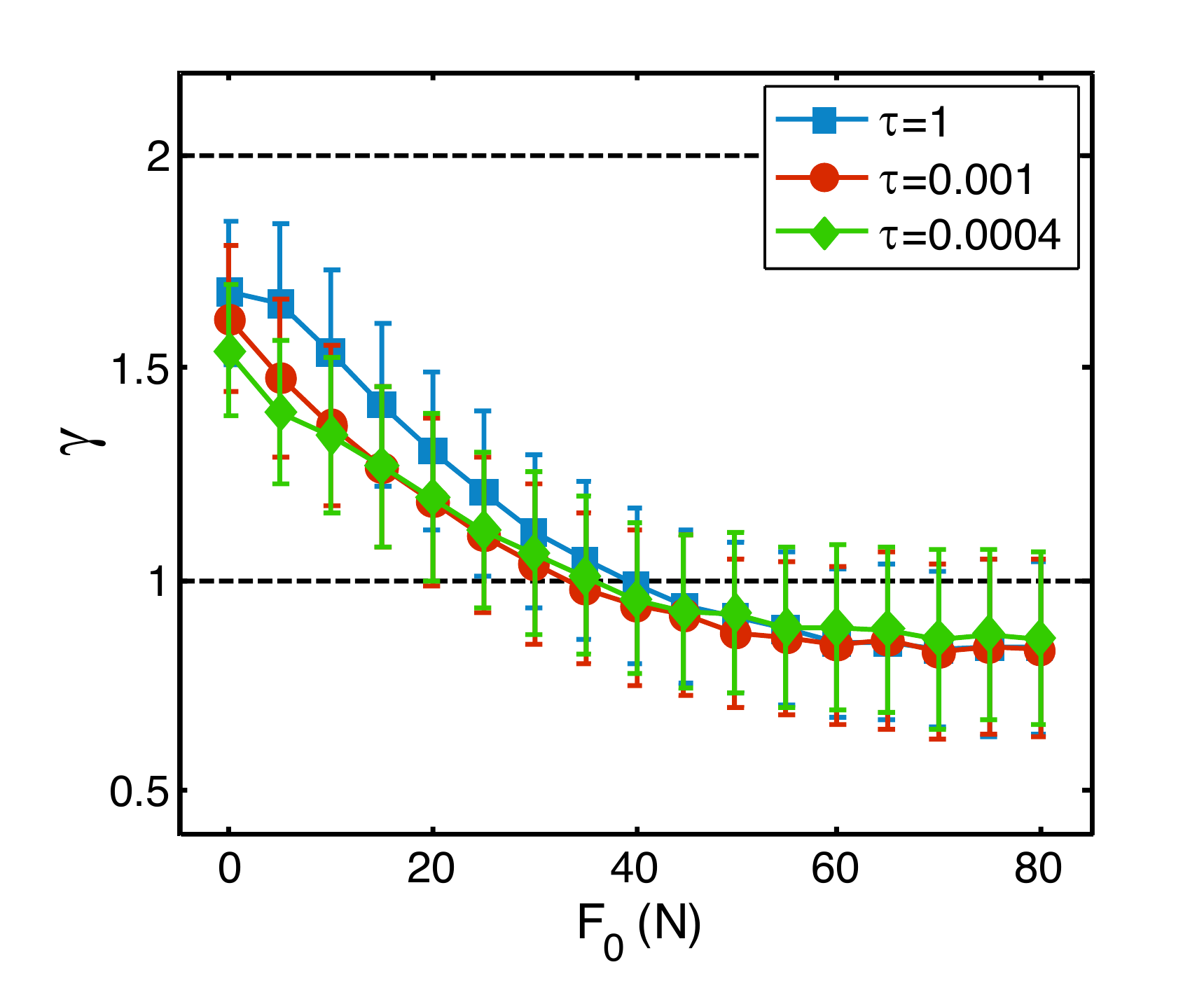}
\caption{\label{fs8} {\bf Effect of dissipation on energy spreading in disordered chains.} Exponent $\gamma$ of the second moment $m_2$ of kinetic energy for various dissipation coefficients ($1/\tau$) as a function of precompression strength.
  We show means of numerical simulations from 100 disordered chains with 100 particles each over the time period $[0.1,3]$ ms.}
\end{figure}

\noindent We investigate the effect of dissipation
on energy spreading. For simplicity, we use a linear damping, as indicated in the last term in Eq.~(3) of the main text. This is a common choice for incorporating dissipative effects in models of granular
chains\cite{ref1}. However, we note in passing that numerous models have been proposed
to capture dissipative effects, and the issue of deriving a proper qualitative and
quantitative incorporation of dissipation is an open problem\cite{chong2016}. 
In Figure~\ref{fs8}, we show the effect of the dissipation time scale
$\tau$ on the exponent $\gamma$ of the second moment $m_2$ of kinetic energy when we vary the static precompression $F_0$.
When the effect of damping is significant ($\tau = 0.0004$), $\gamma$ decreases slightly in the weakly and strongly nonlinear regimes (Compare the blue circles with the red squares and green diamonds in Figure~\ref{fs8}). However, we find that the global dissipation term does not significantly change the characteristics of energy spreading. 

%%%%%%%%%%%%%%%%%%%%%%%%%%%%%%%
%%%%%%%%%%%%%%%%%%%%%%%%%%%%%%%

\end{document}